\documentclass[english,prd,nofootinbib,preprintnumbers]{revtex4}
\usepackage[T1]{fontenc}
\usepackage[latin9]{inputenc}
\usepackage{babel}
\usepackage{amssymb}
\usepackage{graphicx}
\usepackage{esint}
\usepackage[unicode=true,pdfusetitle,
 bookmarks=true,bookmarksnumbered=false,bookmarksopen=false,
 breaklinks=false,pdfborder={0 0 1},backref=false,colorlinks=false]
 {hyperref}
\usepackage{breakurl}

\makeatletter
\@ifundefined{textcolor}{}
{%
 \definecolor{BLACK}{gray}{0}
 \definecolor{WHITE}{gray}{1}
 \definecolor{RED}{rgb}{1,0,0}
 \definecolor{GREEN}{rgb}{0,1,0}
 \definecolor{BLUE}{rgb}{0,0,1}
 \definecolor{CYAN}{cmyk}{1,0,0,0}
 \definecolor{MAGENTA}{cmyk}{0,1,0,0}
 \definecolor{YELLOW}{cmyk}{0,0,1,0}
}

\makeatother

\begin{document}

\title{Nuclear effects in neutrinoproduction of pions}

\author{Iv\'an Schmidt, M.~Siddikov}

\address{Departamento de F\'isica, Universidad Técnica Federico Santa Mar\'ia,\\
 y Centro Cient\'ifico - Tecnol\'ogico de Valpara\'iso, Casilla 110-V, Valpara\'iso,
Chile}

\preprint{USM-TH-333}
\begin{abstract}
In this paper we study nuclear effects in the neutrinoproduction of
pions. We found that in a Bjorken kinematics, for moderate $x_{B}$
accessible in ongoing and forthcoming neutrino experiments, the cross-section
is dominated by the incoherent contribution; the coherent contribution
becomes visible only for small $|t|\lesssim1/R_{A}^{2}$, which requires
$x_{B}\lesssim0.1$. Our results could be relevant to the kinematics
of the ongoing MINERvA experiment in the middle-energy (ME) regime. We provide
a code which could be used for the evaluation of the $\nu$DVMP observables
using different parametrizations of GPDs and different models of nuclear
structure.
\end{abstract}
\maketitle

\section{Introduction}

Today one of the key objects used to parametrize the nonperturbative
structure of the target are the generalized parton distributions (GPDs).
For kinematics where the collinear factorization is applicable~\cite{Ji:1998xh,Collins:1998be},
they allow to evaluate cross-sections for a wide class of processes.
Right now all the information on GPDs comes from electron-proton and
positron-proton measurements done at JLAB and HERA, in particular
the deeply virtual Compton scattering (DVCS) and deeply virtual meson
production (DVMP)~\cite{Ji:1998xh,Collins:1998be,Mueller:1998fv,Ji:1996nm,Ji:1998pc,Radyushkin:1996nd,Radyushkin:1997ki,Radyushkin:2000uy,Collins:1996fb,Brodsky:1994kf,Goeke:2001tz,Diehl:2000xz,Belitsky:2001ns,Diehl:2003ny,Belitsky:2005qn,Kubarovsky:2011zz}.
A planned CLAS12 upgrade at JLAB~\cite{Kubarovsky:2011zz} and ongoing
experiments at COMPASS~\cite{dHose:2014zka} will help to improve
our understanding of the GPDs, and in particular the ability to polarize
both the beam and the target will allow to measure a large number
of polarization asymmetries, providing various constraints for phenomenological
GPD parametrizations. However, in practice the extraction of GPDs
from modern experimental data is still aggravated by uncertainties,
such as large BFKL-type logarithms in next-to-leading order (NLO)
corrections~\cite{Ivanov:2007je} at HERA kinematics, higher-twist
components of GPDs and pion distribution amplitudes (DAs) at JLAB
kinematics~\cite{Ahmad:2008hp,Goloskokov:2009ia,Goloskokov:2011rd,Goldstein:2012az},
and vector meson DAs in the case of $\rho$- and $\phi$-meson production. 

From this point of view, consistency checks of the GPD extraction
from experimental data, especially of their flavor structure, are
important. Earlier we proposed to study the GPDs in deeply virtual
neutrinoproduction of pseudo-Goldstone mesons ($\pi,\, K,\,\eta$)~\cite{Kopeliovich:2012dr}
with high-intensity \textsc{NuMI} beam at Fermilab, which recently
switched to the so-called middle-energy (ME) regime~\cite{Higuera:2014azj},
with an average neutrino energy of about 6~GeV. The $\nu$DVMP measurements
with neutrino and antineutrino beams in this kinematics are complementary
to the electromagnetic DVMP measured at JLAB. In the axial channel,
due to chiral symmetry breaking, we have an octet of pseudo-Goldstone
bosons, which act as a natural probe of the flavor content. Due to
the $V-A$ structure of the charged current, in $\nu$DVMP one can
access simultaneously the unpolarized GPDs, $H,\, E$, and the helicity
flip GPDs, $\tilde{H}$ and $\tilde{E}$. Besides, using chiral symmetry
and assuming closeness of pion and kaon parameters, the full flavor
structure of the GPDs may be extracted. We found~\cite{Kopeliovich:2014pea}
that the higher-twist corrections in neutrino production are much
smaller than in the electroproduction, which gives an additional appeal
to the neutrinoproduction channel.

Unfortunately, in modern neutrino experiments, for various technical
reasons, nuclear targets are much more frequently used than liquid
hydrogen. By analogy with neutrino-induced deep inelastic scattering
($\nu$DIS) on nuclei, one can expect that $\nu$DVMP on nuclei could
be sensitive to many nuclear phenomena such as shadowing, antishadowing,
EMC-effect and Fermi motion. In inclusive processes, all these effects
give contributions of order $\lesssim$10\% in the $0.1\lesssim x_{B}\lesssim0.8$
region relevant for the ongoing and forthcoming $\nu$DVMP experiments~\cite{Amaudruz:1995tq}.
However, there are indications~\cite{Guzey:2005ba} that in the off-forward
kinematics ($t\not=t_{min}$) they could be enhanced. For this reason,
in order to be able to test reliably various GPD models, one should
take into account nuclear effects. We study them in a handbag approach, 
since in the regime of $x_B>0.1$, relevant for the current and forthcoming $\nu$DVMP 
experiments, multiparticle corrections should be negligible.

The paper is organized as follows. In Section~\ref{sec:NuclEffects}
we discuss the framework used for evaluation of nuclear effects. In
Section~\ref{sec:GPDParam} for sake of completeness we list
briefly the parametrizations of GPDs used for our analysis. In Section~\ref{sec:Results}
we present numerical results and draw conclusions.

\section{Nuclear effects}

\label{sec:NuclEffects}There are two types of processes on nuclear
targets, coherent (without nuclear breakup) and incoherent (with nucleus
breakup into fragments). The former contribution is enhanced due to coherence as $\sim A^{2}$, but this effect is relevant only at very small values of $t$:
At larger values of $|t|\sim1/r_{A}^{2},$ where $r_A$ is the nuclear radius, this contribution vanishes rapidly and eventually gets covered by the incoherent contribution.

The typical values of $x_{B}$ accessible in the modern and forthcoming $\nu$DVMP measurements are
$x_{B}\gtrsim0.1$, and for this reason one can neglect multinucleon
coherence effects and describe the process by single-nucleon interactions.
Combining this with the weak binding of the nucleons inside nuclei, we
may use the Impulse Approximation (IA) and write the amplitude of
the process as, 

\begin{equation}
\mathcal{A}_{{\rm coh}}=\int d^{3}\vec{k}\,\rho_{p}\left(\vec{k}-\frac{\vec{\Delta}}{2},\,\vec{k}+\frac{\vec{\Delta}}{2}\right)\mathcal{A}_{p}\left(\vec{k}-\frac{\vec{\Delta}}{2},\,\vec{k}+\frac{\vec{\Delta}}{2},\, q\right)+\int d^{3}\vec{k}\,\rho_{n}\left(\vec{k}-\frac{\vec{\Delta}}{2},\,\vec{k}+\frac{\vec{\Delta}}{2}\right)\mathcal{A}_{n}\left(\vec{k}-\frac{\vec{\Delta}}{2},\,\vec{k}+\frac{\vec{\Delta}}{2},\, q\right),\label{eq:conv}
\end{equation}
where $\vec{k}-\vec{\Delta}/2$ and $\vec{k}+\vec{\Delta}/2$ are
the momenta of the incoming and outgoing nucleons respectively, $\rho_{p}$
and $\rho_{n}$ are the density matrices of the protons and neutrons
inside nuclei, and $\mathcal{A}_{p,n}$ are the amplitudes of the
process on free protons and neutrons~\cite{Guzey:2008fe}. In~(\ref{eq:conv})
we ignore a poorly known and essentially model-dependent contributions
of the so-called non-nucleonic degrees of freedom, which are sometimes
added to the rhs of~(\ref{eq:conv}). Also, we don't include the
contribution of processes in which a final nucleus remains in an excited
isomer $A^{*}$ state: we expect that such processes are suppressed
both at large-$t$ (due to the nuclear formfactor) and small-$t$ (due
to additional factor $\sim t^{n}$ in multipole transitions between
different shells).~

In a Bjorken kinematics region, a collinear factorization theorem
tells us that the scattering amplitude both on the nucleons and nuclei
has a form of the convolution of the GPD of the baryon $H_{A}$ with
a process-dependent hard coefficient function $C(x,\xi)$~%
\footnote{Explicit expressions for the leading twist coefficient functions for
various $\nu$DVMP processes may be found in~\cite{Kopeliovich:2012dr}%
},
\begin{equation}
\mathcal{A}_{{\rm coh}}\sim\int dx\, C\left(x,\,\xi\right)H_{A}\left(x,\,\xi,\, t\right),\label{eq:A_coh}
\end{equation}
which, combined with~(\ref{eq:conv}), yields a convolution relation
for the GPDs of the nucleus~%
\footnote{In what follows we assume for the sake of simplicity that the spin
of the nucleus is zero, which is true for most frequently used nuclear
targets like $^{{\rm 12}}{\rm C}$, $^{40}{\rm Ca}$,$^{40}{\rm Ar}$,$^{56}{\rm Fe}$,
$^{132}{\rm Xe}$. %
},
\begin{equation}
H_{q/A}\left(x,\,\xi,\, t\right)=\int_{0}^{A}\frac{dy}{y}H_{p/A}\left(y,\,\xi,\, t\right)H_{p}\left(\frac{x}{y},\,\frac{\xi}{y},\, t\right)+\int_{0}^{A}\frac{dy}{y}\rho_{n}\left(y,\,\xi,\, t\right)H_{n/A}\left(\frac{x}{y},\,\frac{\xi}{y},\, t\right),\label{eq:H_conv}
\end{equation}
where $y$ is the light-cone fraction of the nuclear momentum carried
by the nucleon, and we introduced the so-called light cone nucleon
distributions $H_{p/A},H_{n/A}$ related to the densities
$\rho_{p,n}$ as
\begin{equation}
H_{i/A}(y,\,\xi,\, t)=m_{N}\int d^{2}k_{\perp}\rho_{i}\left(m_{N}(y+\xi),\vec{k}_{\perp}-\frac{\vec{\Delta}_{\perp}}{2};\, m_{N}(y-\xi),\vec{k}_{\perp}+\frac{\vec{\Delta}_{\perp}}{2}\right),\quad i=p,n.
\end{equation}

\begin{figure}
\includegraphics[scale=0.5]{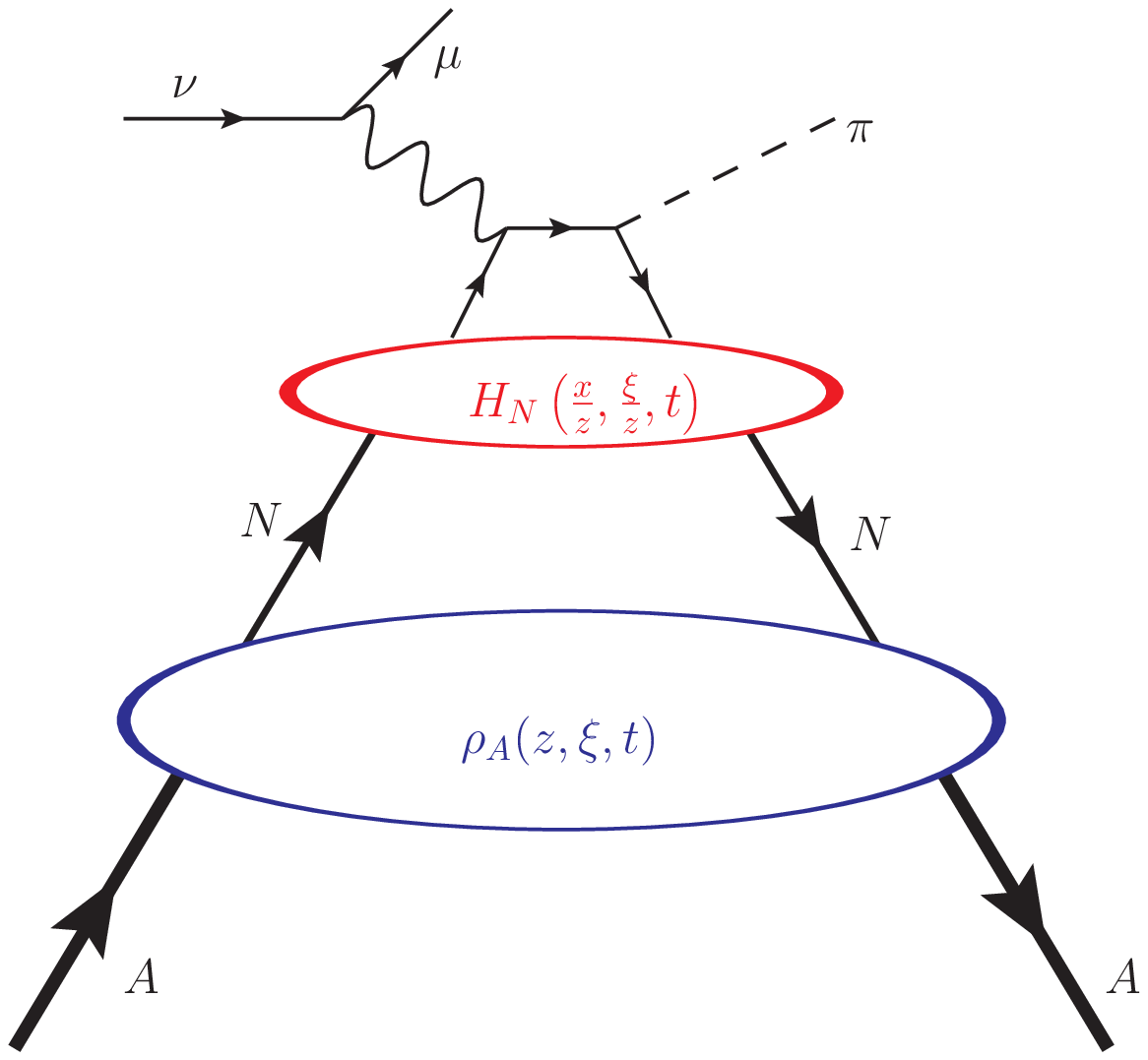}\includegraphics[scale=0.5]{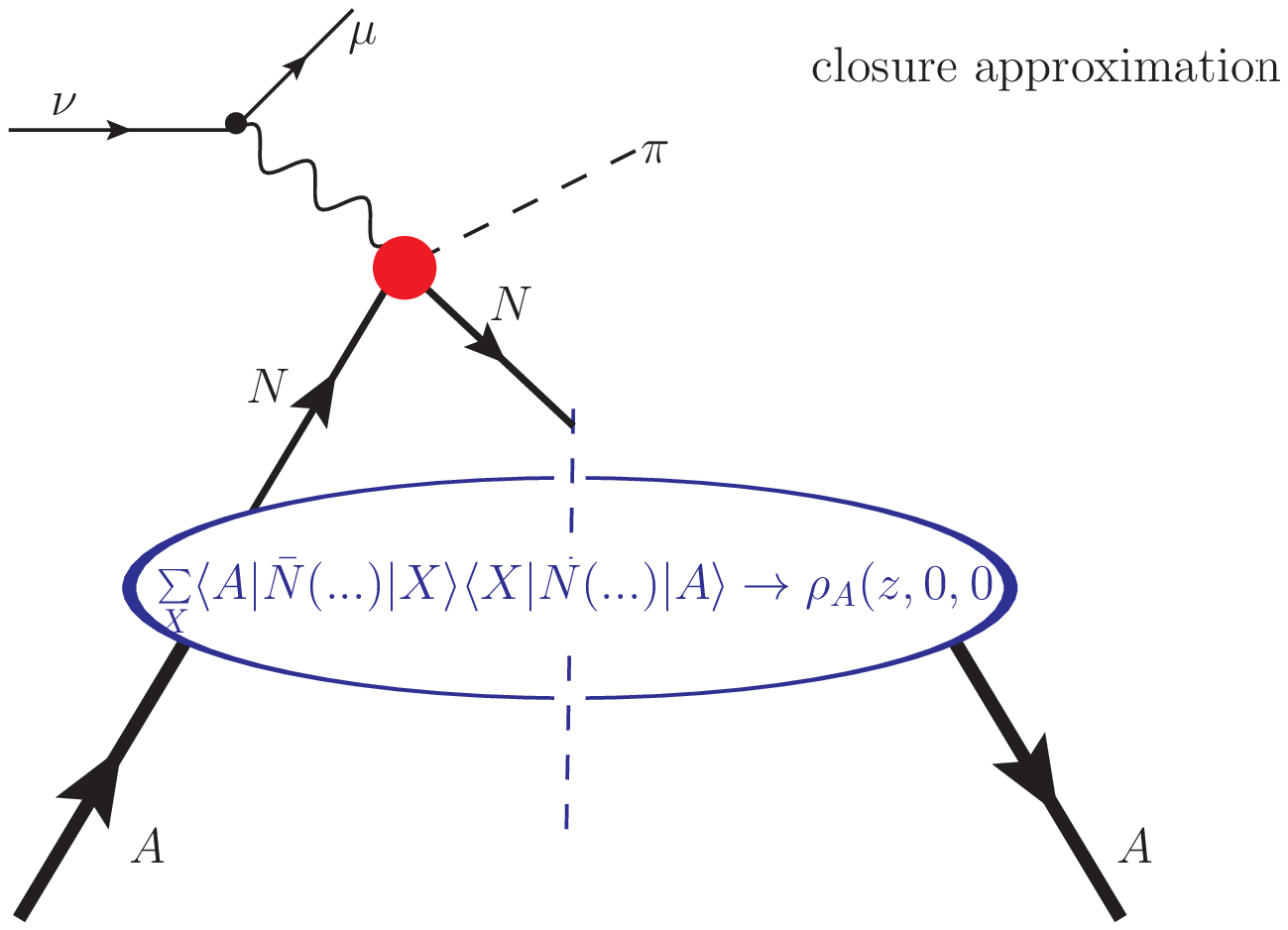}\caption{\textbf{\emph{\label{fig:DoubleHandbag}}}(color online)\textbf{\emph{
Left:}} Double handbag diagram for the amplitude of the coherent pion
production on the nucleus. \textbf{\emph{Right:}} Closure approximation
and its relation to distribution of the cross-section of the incoherent
process. }
\end{figure}

The two equations~(\ref{eq:A_coh},\ref{eq:H_conv}) may be schematically
illustrated with the so-called double handbag diagram in the left
pane of the Figure~\ref{fig:DoubleHandbag}, as a two-stage process.
This approximation has been used e.g. in~\cite{Guzey:2003jh,Liuti:2005gi,Guzey:2005ba,Rinaldi:2014bba,Guzey:2008fe,Schmidt:1977dw},
and describes $e\, A$ data reasonably well. 

For the incoherent processes, we may assume completeness of the final
states (the so-called closure approximation), and using unitarity, as schematically shown by the diagram in the right
pane of the Figure~(\ref{fig:DoubleHandbag}), get a similar expression for the cross-section of the process~\cite{Guzey:2008fe},
\begin{equation}
\sigma_{{\rm incoh}}=\int d^{3}\vec{k}\,\sum_{i=p,n}\rho_{i}\left(\vec{k},\,\vec{k}\right)\sigma_{i}\left(\vec{k},\,\vec{k}+\vec{\Delta},\, q\right)
\approx \int \frac{dy}{y}\,\sum_{i=p,n}H_{i/A}(y,0,0)\sigma_{i}\left(\vec{k},\,\vec{k}+\vec{\Delta},\, q\right)_{\vec k=y \vec{P_A}/A},
\label{H_A_incoh}
\end{equation}
where $P_A$ is the momentum of the nucleus, and the last equality in (\ref{H_A_incoh}) is valid in a collinear approximation.

Since the binding energy of a nucleon in the nucleus is very small
compared to a mass of the free nucleon, the distributions $H_{p/A},\,H_{n/A}$
are strongly peaked functions, and in the first approximation may be approximated as~\cite{Guzey:2005ba}
\begin{eqnarray}
H_{p/A}\left(y,\,\xi\approx0,\, t\approx0\right) & = & Z\,\sqrt{\frac{\alpha}{\pi}}\exp\left[-\alpha\left(y-1\right)^{2}\right]\frac{F_{A}(t)}{F_{A}(0)},\label{eq:rho_p_gauss}\\
H_{n/A}\left(y,\,\xi\approx0,\, t\approx0\right) & = & \left(A-Z\right)\,\sqrt{\frac{\alpha}{\pi}}\exp\left[-\alpha\left(y-1\right)^{2}\right]\frac{F_{A}(t)}{F_{A}(0)},\label{eq:rho_n_gauss}
\end{eqnarray}
where $Z$ is the atomic number, $A$ is the mass number, and the
parameter $\alpha\approx k_{{\rm F}}^{2}/m_{N}^{2}\approx200$~MeV
is the Fermi momentum inside the nucleus, which controls the width of the
distribution. In the extreme limit $\alpha\to0$, the product of the exponent
in~(\ref{eq:rho_p_gauss},\ref{eq:rho_n_gauss}) and a prefactor
$\sqrt{\alpha/\pi}$ reduce to a $\delta$-function, and instead of
a convolution we end up with a mere sum of amplitudes (for the coherent
case) or cross-sections (for the incoherent case) on separate nucleons.
However, such a factorized form is an oversimplification, since it
cannot describe the $A$-dependence of the first moment of the so-called
$D$ -term $d_{A}(0)$, for which there are estimates based on very
general assumptions~\cite{Polyakov:2002yz}. 

A more realistic approach is to use the functions $H_{p/A},\, H_{n/A}$,
evaluated in the shell model of the nuclear structure. One of the
most popular choices for the evaluation of the nucleon dynamics inside
a nucleus is a QHD-I model proposed in~\cite{ChinWalecka:1974,SerotWalecka:1986,SerotWalecka:1997}.
The lagrangian of this model, in its simplest form describes the interaction
of the nucleons with effective vector and scalar fields, 
\begin{equation}
\mathcal{L}=\bar{\psi}\left(i\hat{\partial}-M-g_{v}\hat{V}+g_{s}\phi\right)\psi+\frac{1}{2}\left(\partial_{\mu}\phi\partial^{\mu}\phi\right)-\frac{1}{4}V_{\mu\nu}V^{\mu\nu}+\frac{m_{V}^{2}}{2}V_{\mu}V^{\mu}.\label{eq:L_QHD}
\end{equation}
where we used a shorthand notation $V_{\mu\nu}=\partial_{[\mu}V_{\nu]}$
, $V_{\mu}$ and $\phi$ are the fields of vector and scalar mesons
respectively. The mean field models based on a Lagrangian of type
(\ref{eq:L_QHD}), have been successful in the description of various
characteristics of nuclei. The simplest version of the model used
in this work consists of baryons and isoscalar scalar and vector mesons.
The pseudoscalar pion degrees of freedom are neglected because their
contribution to the ground state of $0^{+}$-nuclei essentially averages
to zero~\cite{SerotWalecka:1986}. In the literature one may find
extensions of the model~(\ref{eq:L_QHD}), which have additional mesonic
degrees of freedom and give better quantitative description of 
nuclei, especially with nonzero spin and isospin. The corresponding
explicit expression for the distribution functions $H_{p/A}$, $H_{n/A}$
were calculated in the model~(\ref{eq:L_QHD}) in~\cite{Guzey:2005ba},
yielding
\begin{equation}
H_{p,n/A}\left(y,\xi,t\right)=\sum_{i}\int\frac{d^{2}k_{\perp}}{\left(2\pi\right)^{2}}\Phi_{i}^{\dagger}\left(y-\xi,\vec{k}_{\perp}+\frac{\vec{\Delta}_{\perp}}{2}\right)\gamma_{+}\Phi_{i}\left(y+\xi,\vec{k}_{\perp}-\frac{\vec{\Delta}_{\perp}}{2}\right),\label{eq:H_A_Shell}
\end{equation}
where $\Phi_{i}$ is the wave function of the nucleon inside the nucleus,
the summation index $i$ runs over the proton or neutron shells respectively.

\section{GPD Parametrization}

\label{sec:GPDParam}For numerical estimates of the nuclear effects,
one should use a particular parametrization of GPDs available from
the literature~\cite{Diehl:2000xz,Goloskokov:2008ib,Radyushkin:1997ki,Kumericki:2011rz,Guidal:2010de,Polyakov:2008aa,Polyakov:2002wz,Freund:2002qf,Goldstein:2013gra}.
For the sake of definiteness, in what follows we use the parametrization
of Kroll-Goloskokov~\cite{Goloskokov:2006hr,Goloskokov:2007nt,Goloskokov:2008ib},
which succeeded to describe HERA~\cite{Aaron:2009xp} and JLAB~\cite{Goloskokov:2006hr,Goloskokov:2007nt,Goloskokov:2008ib}
data on electroproduction of different mesons, and therefore it should provide
a reasonable description of neutrino-induced DVMP. The parametrization
is based on the Radyushkin's double distribution ansatz, in which
the skewness is introduced separately for sea and valence quarks,
\begin{equation}
H(x,\xi,t)=H_{val}(x,\xi,t)+H_{sea}(x,\xi,t),
\end{equation}
where 
\begin{eqnarray}
H_{val}^{q} & = & \int_{|\alpha|+|\beta|\le1}d\beta d\alpha\delta\left(\beta-x+\alpha\xi\right)\,\frac{3\theta(\beta)\left((1-|\beta|)^{2}-\alpha^{2}\right)}{4(1-|\beta|)^{3}}q_{val}(\beta)e^{\left(b_{i}-\alpha_{i}\ln|\beta|\right)t},\label{eq:H_val}\\
H_{sea}^{q} & = & \int_{|\alpha|+|\beta|\le1}d\beta d\alpha\delta\left(\beta-x+\alpha\xi\right)\,\frac{3\, sgn(\beta)\left((1-|\beta|)^{2}-\alpha^{2}\right)^{2}}{8(1-|\beta|)^{5}}q_{sea}(\beta)e^{\left(b_{i}-\alpha_{i}\ln|\beta|\right)t},\label{eq:H_sea}
\end{eqnarray}
and $q_{val}$ and $q_{sea}$ are the ordinary valence and sea components
of PDFs. The coefficients $b_{i}$, $\alpha_{i}$, as well as the
parametrization of the input PDFs $q(x),\,\Delta q(x)$ and pseudo-PDFs
$e(x),\,\tilde{e}(x)$ (which correspond to the forward limit of the
GPDs $E,\,\tilde{E}$) are discussed in~\cite{Goloskokov:2006hr,Goloskokov:2007nt,Goloskokov:2008ib}.
The unpolarized PDFs $q(x)$ are adjusted to reproduce the CTEQ PDFs
in the limited range $4\lesssim Q^{2}\lesssim40$~GeV$^{2}$. The
$\nu$DVMP cross-sections on free protons and neutrons have been evaluated
with this parametrization in our previous papers~\cite{Kopeliovich:2012dr,Kopeliovich:2014pea}.

\section{Results and discussion}

\label{sec:Results}In order to quantify the size of the nuclear effects,
we consider a ratio 

\begin{equation}
R_{A}=\frac{d\sigma_{A}/dt\, d\nu\, dQ^{2}}{Z\, d\sigma_{p}/dt\, d\nu\, dQ^{2}+\left(A-Z\right)\, d\sigma_{n}/dt\, d\nu\, dQ^{2}}.\label{eq:R_A-1}
\end{equation}
which takes into account differences in isotopic content of different
nuclei. For the case of self-conjugate nuclei, this ratio up to a
coefficient $A/2$ coincides with a deuteron-normalized cross-section~%
\footnote{The nuclear effects in the deuteron are small.  The shadowing corrections
are also negligible since we consider the kinematics $x_{B}\gtrsim0.1$.%
} used in the presentation of experimental data. 

\begin{figure}
\includegraphics[scale=0.4]{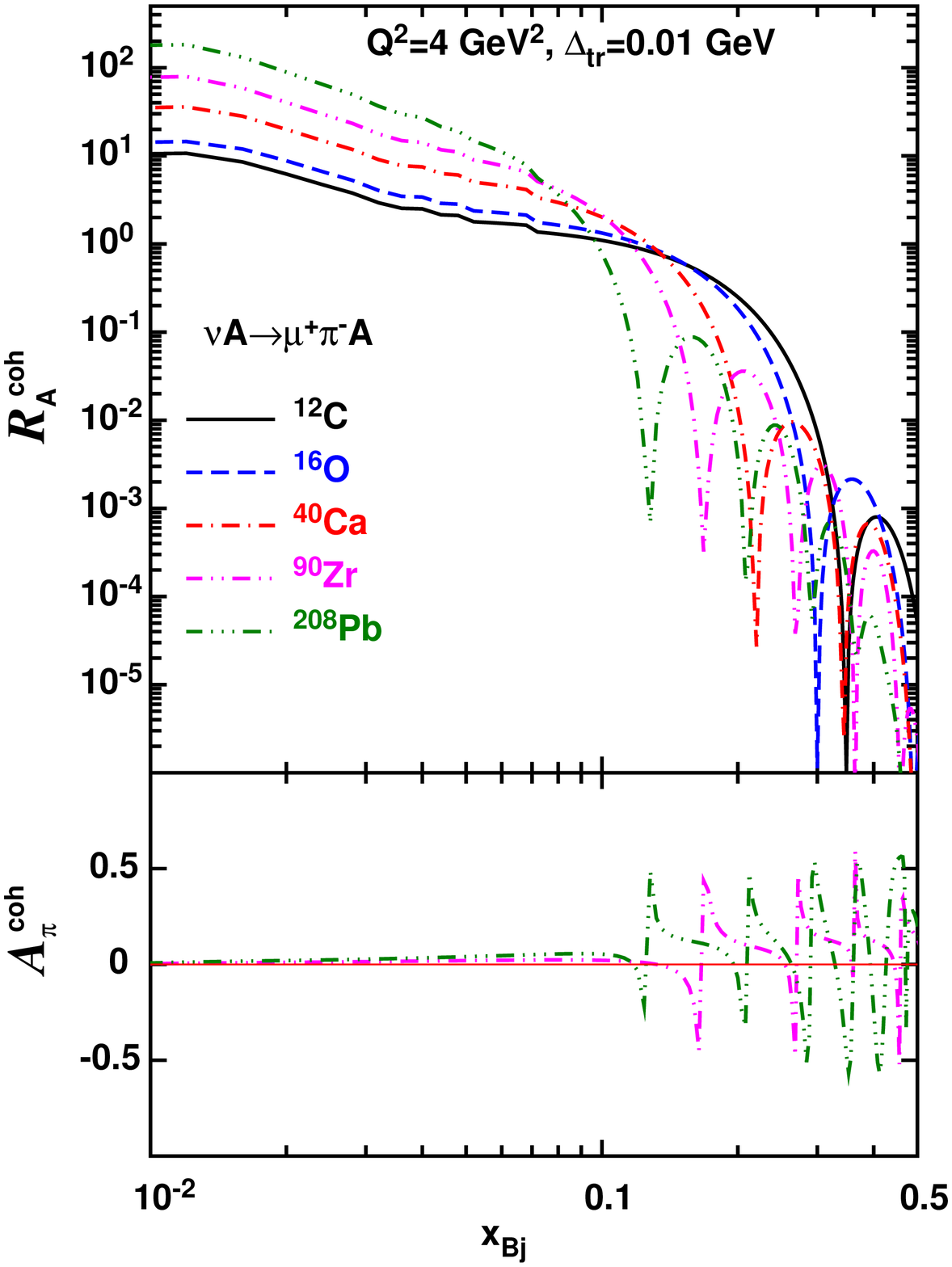}\includegraphics[scale=0.4]{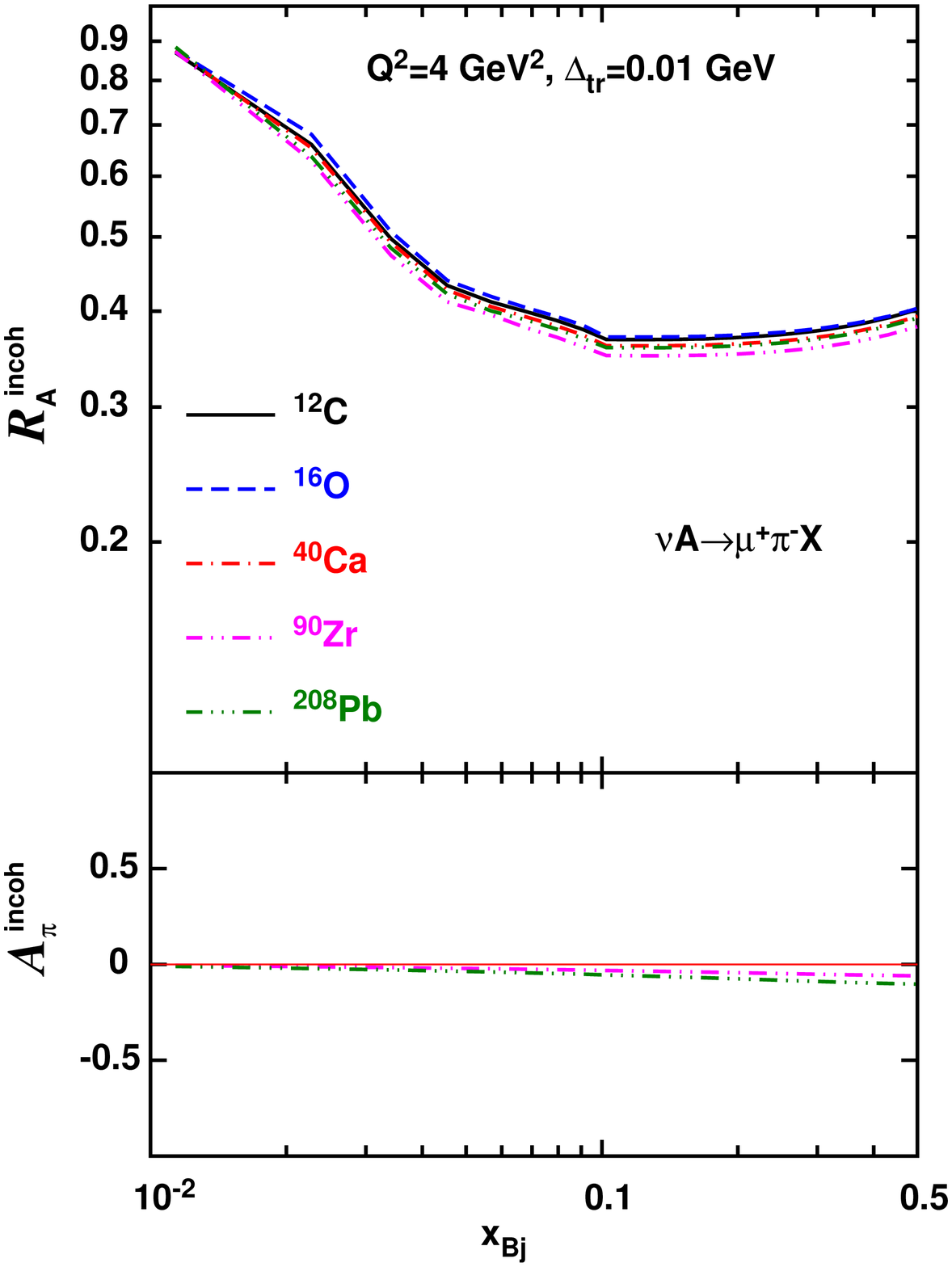}

\caption{\label{fig:coherent_Ratio-x}(color online) Ratio~(\ref{eq:R_A-1})
for coherent (left) and incoherent (right) $\pi^{-}$-production for
several nuclei. In the lower parts of each figure we show the asymmetry~(\ref{eq:Asym}),
which measures the difference between $\pi^{+}$ and $\pi^{-}$ production.
As explained in the text, for the first three nuclei it is exactly
zero, so for the sake of legibility we don't show those curves.}
\end{figure}

In the left pane of Figure~(\ref{fig:coherent_Ratio-x}), we have
plotted the ratio $R_{A}$~(\ref{eq:R_A-1}) for several spin-0 nuclei.
In the regime of small-$x_{B}$ the ratio $R_{A}$ is close to $A$
due to the coherence of contributions of separate nucleons. For higher
values of $x_{B}$, the behavior of the cross-section is similar
to that of a formfactor: it decreases rapidly and has nodes, with
average distance between the nodes $\sim1/r_{A}$, where $r_{A}$
is the nuclear radius. However, the positions of the nodes do not
coincide with those of a nuclear formfactor, a type of behavior which
cannot be reproduced by a simple model~(\ref{eq:rho_p_gauss},\ref{eq:rho_n_gauss}).
In neutrino experiments this kinematic region is hardly accessible
experimentally, since it is covered by the incoherent contribution if the final nucleus breakup is not detected (see the right pane of the same Figure). In the
lower pane of each figure, we've shown the asymmetry 
\begin{equation}
A_{\pi}=\frac{d\sigma_{\nu A\to\mu^{+}\pi^{-}A}-d\sigma_{\bar{\nu}A\to\mu^{-}\pi^{+}A}}{d\sigma_{\nu A\to\mu^{+}\pi^{-}A}+d\sigma_{\bar{\nu}A\to\mu^{-}\pi^{+}A}}\sim\int d^{3}k\,\left(\rho_{p}(k,\, k)-\rho_{n}(k,\, k)\right)\left(d\sigma_{\nu p\to\mu^{+}\pi^{-}p}-d\sigma_{\nu n\to\mu^{+}\pi^{-}n}\right),\label{eq:Asym}
\end{equation}
which is sensitive to an isospin-1 GPD combination $H^{u}-H^{d}$.
For self-conjugate nuclei, this asymmetry is exactly zero since in
the model~\cite{SerotWalecka:1986,SerotWalecka:1997} the difference
between proton and neutron distributions is negligible. For $^{90}$Zr
and $^{208}$Pb, the asymmetry~(\ref{eq:Asym}) in general is small
and does not exceed 10\%, although increases slightly near the nodes
of $\pi^{+}$ and $\pi^{-}$~%
\footnote{The nodes of $\pi^{+}$ and $\pi^{-}$ don't exactly match due to
differences in proton and neutron distributions $\rho_{p},\,\rho_{n}$.%
}.

\begin{figure}
\includegraphics[scale=0.4]{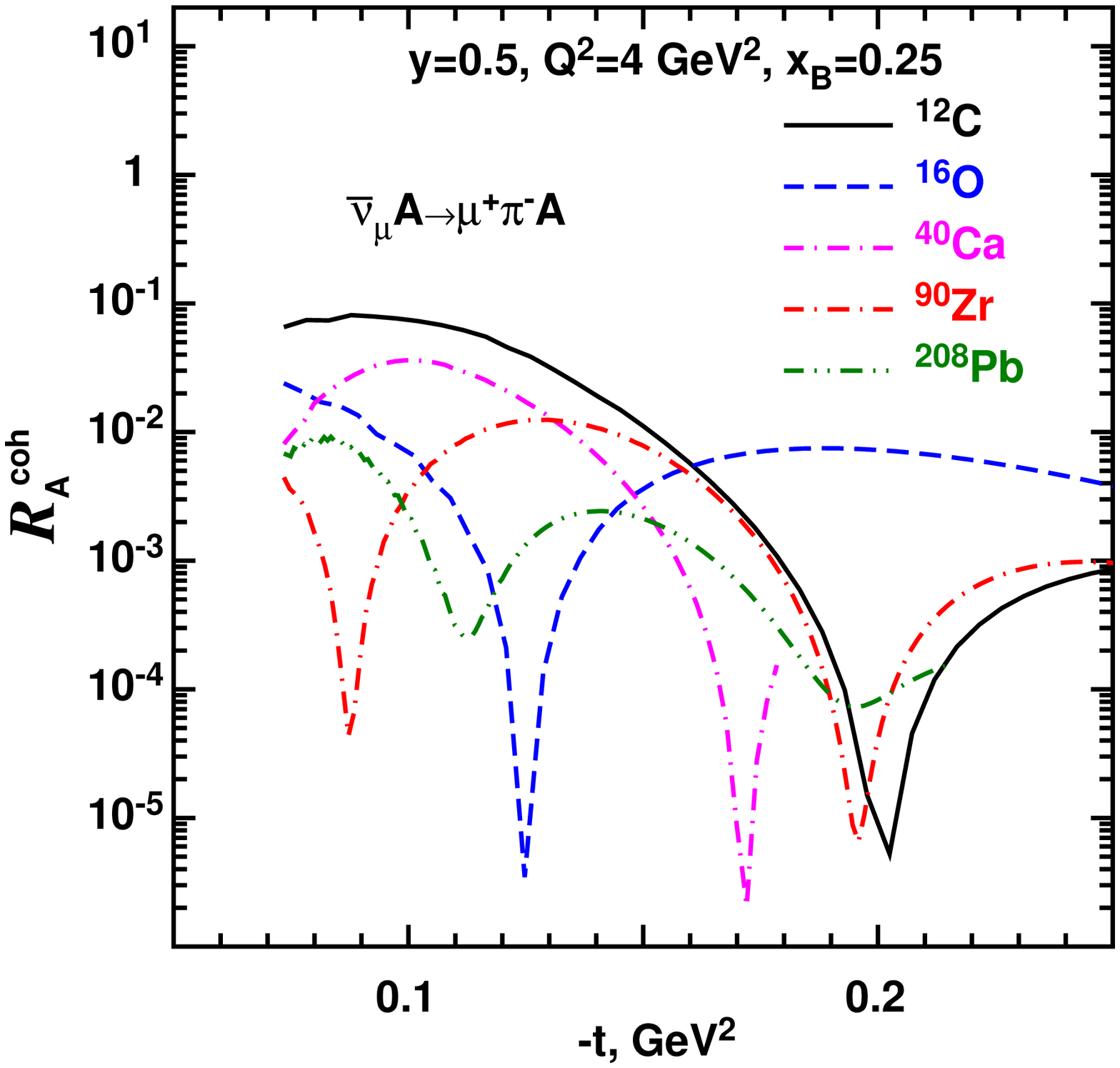}\includegraphics[scale=0.4]{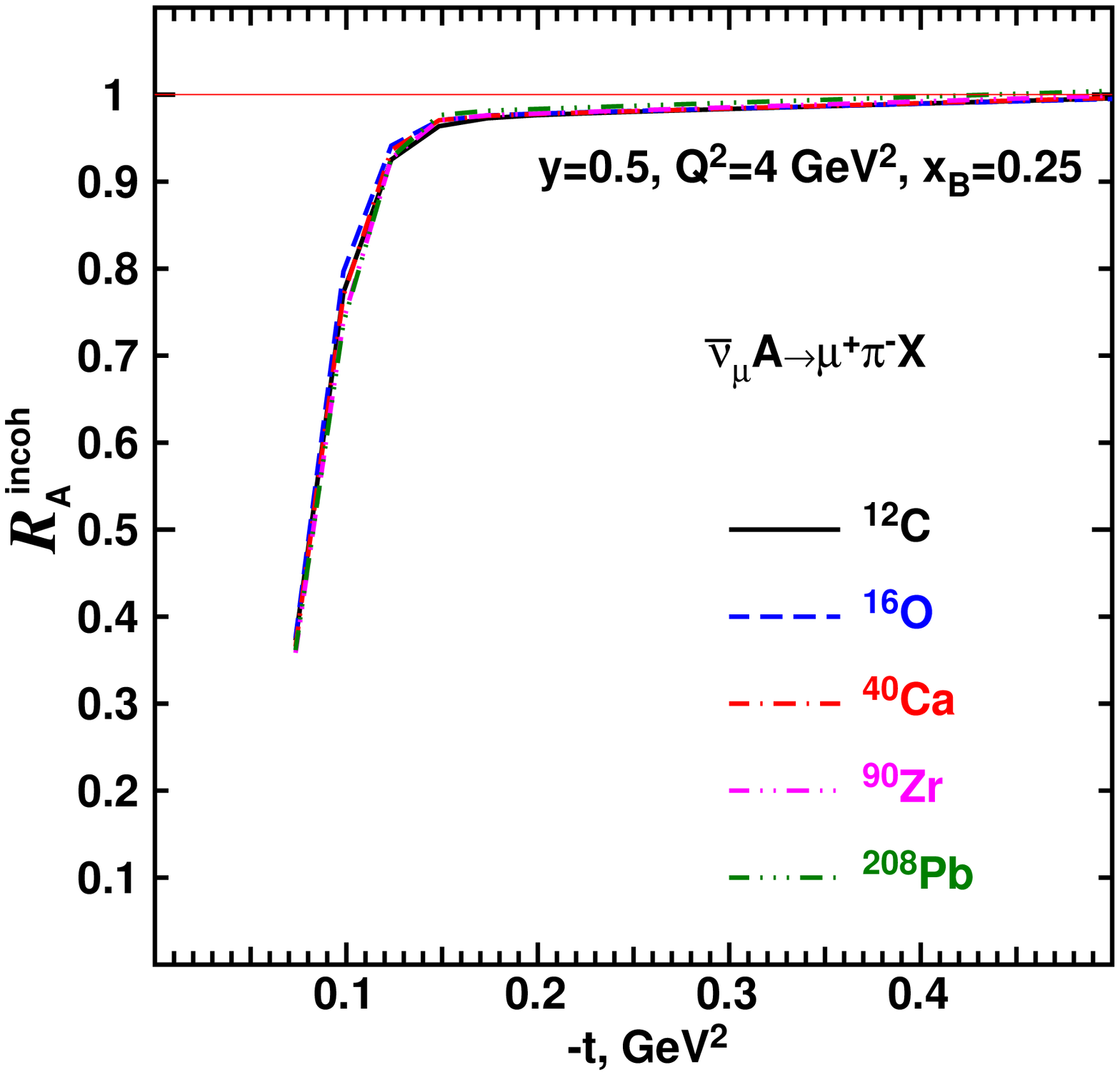}

\caption{\label{fig:coherent_Ratio}(color online) $t$-dependence of the ratio~(\ref{eq:R_A-1})
for coherent (left) and incoherent (right) $\pi^{-}$-production for
several nuclei.}
\end{figure}

The $t$-dependence of the cross-section is shown in Figure~\ref{fig:coherent_Ratio}.
At small-$t$, the coherent ratio $R_{A}$ scales as $R_{A}\propto A\,\exp(t_{min}(x_{B})r_{A}^{2}/6)$,
where $r_{A}$ is the nuclear radius, but decreases rapidly at large-$t$.
The incoherent cross-section shown schematically in the right pane
of Figure~\ref{fig:coherent_Ratio},  is close to unity, as expected.
Its suppression at small-$|t|\sim|t_{min}|$ comes from $\xi_{p}\le1$
and onshellness conditions in the convolution integral in~(\ref{eq:H_conv}).

\begin{figure}
\includegraphics[scale=0.4]{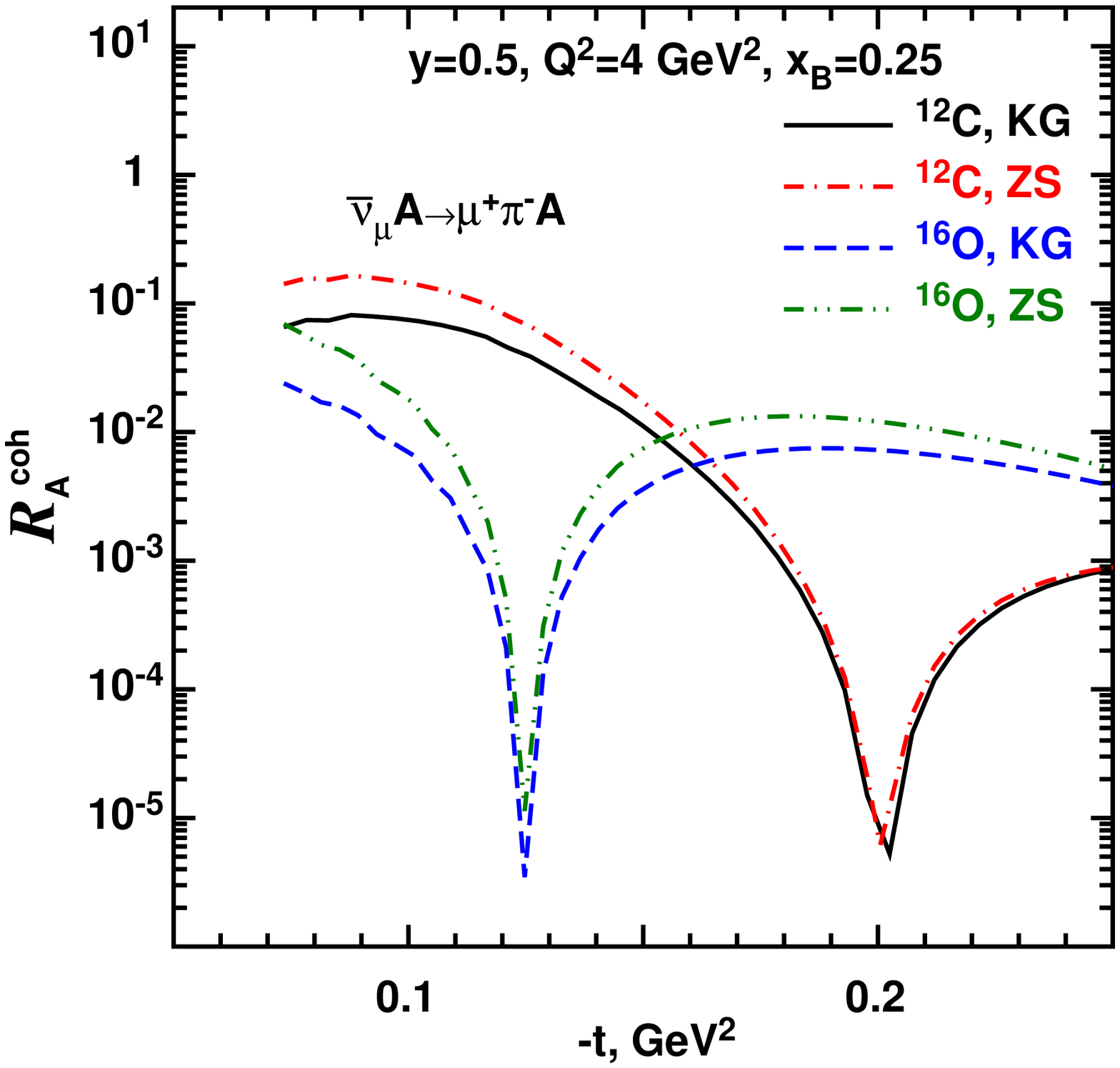}
\includegraphics[scale=0.4]{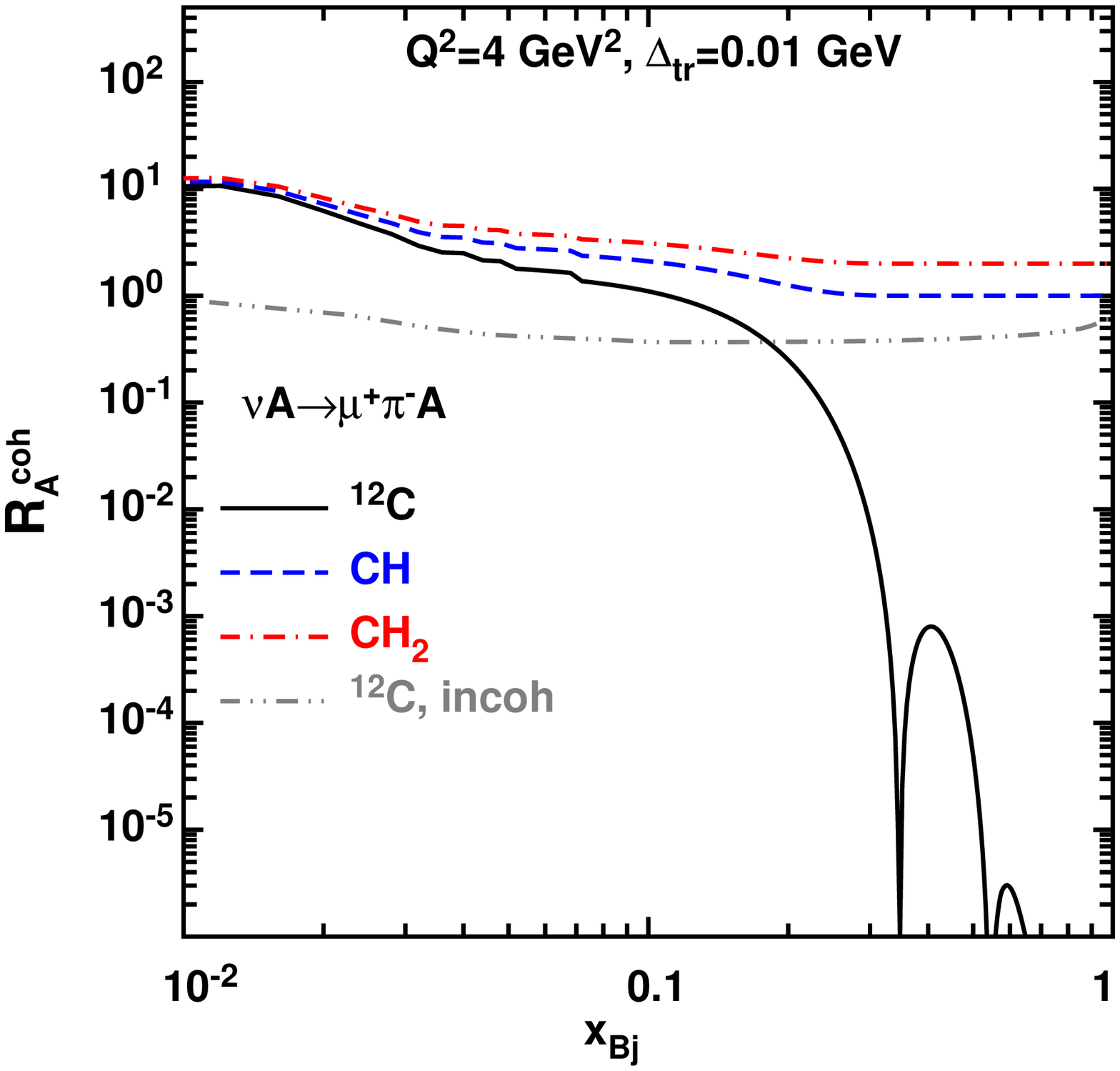}

\caption{\label{fig:coherent_Ratio-atoms}(color online) Left: Comparison of coherent cross-sections evaluated in Kroll-Goloskokov (KG) and Zero Skewness (ZS) models. Right: Comparison of different contributions to pion production on atomic targets $\mathbf{CH_{n}}.$ }
\end{figure}

In order to understand the sensitivity of the cross-sections to a choice of GPD parametrization, in the left pane of the Figure~\ref{fig:coherent_Ratio-atoms} we compared the predictions of Kroll-Goloskokov model discussed in Section~\ref{sec:GPDParam} with a simple zero-skewness model $H^{q}=q(x)F_N(t)$. As we can see, there is an up to a factor of two difference between the two models.

Frequently the targets in neutrino experiments are organic scintillators
with a general atomic structure $\mathbf{CH_{n}}$. As one can see
from the right pane of the Figure~\ref{fig:coherent_Ratio-atoms}, in the region $x_{B}\gtrsim0.3$
there are two dominant contributions, from hydrogen atoms and from
incoherent cross-sections, which cannot be separated unless a final
nucleus is detected. A coherent cross-section is strongly suppressed
in this kinematics.

\section{Conclusions}

In this paper we studied the nuclear effects in the coherent and incoherent
pion production. We found that the former has a complicated structure,
with coherent enhancement in the region of small-$x_{B}$, small-$t$,
and strong nuclear suppression outside this kinematics. Similar to
a formfactor, the leading twist contribution has nodes. For the incoherent
case, the nuclear dependence is quite mild, and for $|t|\gtrsim3|t_{min}(x_{B})|$
the nuclear effects are negligible, i.e the full cross-section is
a mere sum of contributions of separate nucleons. This is a model-independent
result, and from a practical point of view, this allows to get rid
of extra uncertainties related to nuclear structure. Our approach is 
applicable in the regime $x_{B}\gtrsim10^{-2}$. For smaller values of $x_{B}$,
this picture is modified due to coherence~\cite{Kopeliovich:2012kw} and saturation~\cite{Rezaeian:2012ji} effects. For further
practical applications, we provide a code, which can be used for the
evaluation of nuclear cross-sections with different parametrizations
of GPDs and models of nuclear structure. The modular structure of the code allows to easily consider different parametrizations of GPDs and nuclear distributions.
For illustration, we provide with this package the libraries for the Kroll-Goloskokov GPD model and the QHD-I nuclear structure model distribution used in this paper. 
Also, we provide detailed instructions how to build and use new libraries.

Finally, we would like to stop briefly on the recent results of the
MINERvA collaboration~\cite{Higuera:2014azj}. Albeit those results
are for coherent pion production on nuclei, here we do not make any
comparison, because the conditions of applicability of collinear factorization
($Q^{2}\gg m_{N}^{2}$) are not met in that kinematics. Models based
on extrapolation of the Adler relation~\cite{Adler:1964yx} are more
appropriate for that kinematics.

\section*{Acknowledgments}

This work was supported in part by Fondecyt (Chile) grants No. 1140390
and 1140377.

\appendix

  \end{document}